\documentclass[useAMS,usenatbib]{mn2e}
\usepackage{psfig,subfigure}
\title[Modeling the Photopolarimetric Variability of AA Tau]
 {Modeling the Photopolarimetric Variability of AA Tau}
\author[O'Sullivan, et al.]
{{Mark O'Sullivan$^1$, Michael Truss$^1$, Christina Walker$^1$, 
Kenneth Wood$^1$, Owen Matthews$^2$} 
\newauthor
{Barbara Whitney$^3$, \& J.E.~Bjorkman$^4$} \\
$^1$School of Physics \& Astronomy, University of St. Andrews,\\
North Haugh, St Andrews, KY16 9SS, Scotland\\
$^2$Laboratory for Astrophysics, Paul Scherrer Institute, 
Wurenlingen und Villigen, CH-5232, Villigen PSI, Switzerland\\
$^3$Space Science Institute, 3100 Marine Street, Suite A353, 
Boulder, CO~80303\\
$^4$Ritter Observatory, Department of Physics \& Astronomy, 
University of Toledo, Toledo, OH 43606}

\date{Released 2004 Xxxxx XX}
\pagerange{\pageref{firstpage}--\pageref{lastpage}} \pubyear{2004}

\def\LaTeX{L\kern-.36em\raise.3ex\hbox{a}\kern-.15em
 T\kern-.1667em\lower.7ex\hbox{E}\kern-.125emX}

\begin{document}

\label{firstpage}

\maketitle

\begin{abstract}

We present Monte Carlo scattered light models of a warped disk that 
reproduce the observed photopolarimetric variability of the classical 
T~Tauri star AA~Tau.  For a system inclination of $75^\circ$ and using an 
analytic description for a warped inner disk, we 
find that the shape and amplitude of the photopolarimetric variability 
are reproduced with a warp which occults the star, located at 0.07~au, 
amplitude 0.016~au, extending over radial and azimuthal ranges $0.0084$~au 
and $145^\circ$. We also show a time sequence of high spatial resolution 
scattered light images, showing a dark shadow cast by the warp sweeping round 
the disk. Using a modified smooth particle hydrodynamics code, we find that a 
stellar dipole magnetic field of strength 5.2~kG, inclined at $30^\circ$ to 
the stellar rotation axis can reproduce the required disk warping to explain 
AA~Tau's photopolarimetric variability.  

\end{abstract}

\begin{keywords} stars: circumstellar matter --- individual: AA Tau

\end{keywords}

\section{Introduction}

Observationally, pre-main-sequence T~Tauri stars may be divided into 
classical and weak T~Tauri stars (Appenzeller \& Mundt 1989; Bertout 
1989). The classical T~Tauri stars (cTTS) exhibit broad H$\alpha$ indicative 
of on-going accretion, and are surrounded by large dusty disks inferred from 
their signature infrared excess emission.  On the other hand, weak T~Tauri 
stars (wTTS) do not exhibit the strong H$\alpha$ and large IR excesses and 
these systems may have ceased accreting and have very low mass disks. Both 
cTTS and wTTS exhibit photometric variability (a defining feature of T~Tauri 
stars, Joy 1945), with wTTS exhibiting periodic and cTTS quasi-periodic 
variability (Appenzeller \& Mundt 1989; Bertout 1989). Hot and cool spots 
on the stellar surface are believed to be responsible for the variability, 
with long lived cool spots dominant in wTTS (Hatzes 1995) and short 
lived hot spots, possibly linked to the accretion process, dominating in 
cTTS (Kenyon et al. 1994; Bouvier et al. 1993; Eaton,Herbst \& 
Hillenbrand 1995; Choi \& Herbst 1996; Herbst et al. 1994). The currently 
popular magnetospheric accretion model predicts hot spots on the stellar 
surface (Ghosh \& Lamb 1979; Konigl 1991; Shu et al. 1994).  In this model a 
stellar dipole magnetic field threads, truncates, and possibly warps the 
circumstellar disk.  Disk material is accreted onto the star along magnetic 
field lines forming hot spots or rings on the stellar surface at the magnetic 
poles (Mhadavi \& Kenyon 1998).  

The magnetospheric accretion model has been applied to explain the 
photometric variability in the cTTS system DR Tau (Kenyon et al. 1994). For 
the edge-on disk around HH~30~IRS, variability has been observed in both HST 
images (Burrows et al. 1996; Stapelfeldt et al. 1999; Cotera et al. 2001) and 
ground based $VRI$ photometry (Wood et al. 2000). The variability in HH~30~IRS 
has been modelled as due to hot spots (Wood \& Whitney 1998, Wood et al. 
2000) and a disk warp (Stapelfeldt et al. 1999), but no period has so far 
been determined. Photopolarimetric variability has recently been observed 
in the cTTS AA~Tau (Bouvier et al. 1999, 2003) and has been interpreted 
as due to occultation of a star by a warp in the inner regions of a 
circumstellar disk. The variability is found to have a period of 8.2~days, 
but the variations sometimes turn-off, possibly due to non-steady accretion.  

In this paper, we extend our Monte Carlo photopolarimetry simulations 
of systems with hot star spots (Wood et al. 1996; Wood \& Whitney 1998; 
Stassun \& Wood 1999) to include the effects of disk warps. We construct 
a warped disk model that reproduces the photopolarimetric variability of 
AA~Tau. Then, using a modified SPH code incorporating forces from an 
inclined dipole stellar magnetic field, we explore the magnetic field 
strength and configuration required to warp the AA~Tau disk and reproduce 
the observed photopolarimetric variability.  \S~2 describes the 
photopolarimetric data. \S~3 presents models for the system using the SED to 
constrain the large scale disk structure, photpolarimetric variability to 
constrain the warping of the inner disk, and we also present a time sequence 
of high spatial resolution scattered light images. \S~4 presents our SPH 
simulations of a disk with an inclined stellar dipole field, and we summarise 
our results in \S~5.

\section{AA~Tau Photopolarimetry}

\setcounter{enumi}{5}

The photopolarimetry of AA Tau has been reported in a number of studies. 
The young star at the heart of the system has been classified as a 
K7$\Roman{enumi}$ (Kenyon \& Hartmann 1995), with a mass of $0.8M_\odot$, 
radius of $1.85R_\odot$ and an effective temperature of $4030\pm30$~K 
(Bouvier et al. 1999). From analysis of the photometric variability the 
system's disk, inferred from the IR excess emission 
(e.g., D'Alessio et al. 1999), is estimated to be at an inclination of 
$70^\circ$ or greater (Bouvier et al. 1999). The observed photometric 
variability is achromatic and is attributed to occultation of the star by a 
warp in the inner disk (Terquem \& Papaloizou 2000; Bouvier et al. 2003). The 
occultation results in $\Delta V\sim 1$~mag, has a duration of around 
3-4~days, and recurs every 8.2~days, but occasionally an occultation event is 
missing. Observations show the overall brightness level and depth of 
eclipses are variable (Bouvier et al. 2003),  suggestive of stochastic 
magnetospheric accretion in AA~Tau. Throughout this study the 'time averaged'
photopolarimetric variations given here are used as the basis for the model. 
Assuming a Keplerian disk, the warp responsible for the periodic occultation 
must be located at 0.07~au (Bouvier et al. 2003).

Polarimetry studies show that the linear polarisation increases as the 
observed flux decreases and that the polarisation has a range of $0.6\% - 
1.3\%$ (M\'{e}nard et al. 2003). The polarisation position angle is shown to 
vary from $0^\circ$ to almost $30^\circ$ (Bouvier et al. 1999) and studies 
of nearby stars exhibit similarly large position angle variations.  
The large position angle variations are attributed to interstellar 
polarisation of around 0.5\%, so the intrinsic polarization is variable 
in the range $\sim$ $0.1\% - 0.8\%$ (M\'{e}nard et al. 2003).

\section{Radiation Transfer Models}

We model AA~Tau's SED and photopolarimetry using our suite of Monte Carlo 
scattered light and radiative equilibrium codes.  For the SED models we 
use an axisymmetric disk to determine the disk shape and mass.  The 
photopolarimetric modelling uses a non-axisymmetric warped inner disk 
described below.

As we model the time dependent photopoalrimetry with a warped disk,
a fully self-consistent model should calculate the 3D time-dependent disk
temperature and density structure and time dependent SED. Such a calculation 
is beyond the scope of this paper and instead we present three separate 
models. The first models AA~Tau's SED with an axisymmetric disk. The disk 
structure is calculated by enforcing vertical hydrostatic equilibrium in the 
disk as described in Walker et al.(2004). In the second model we model the 
photopolarimetry using an analytic description for warping of the inner disk 
and for the outer disk we use the hydrostatic disk structure derived from the 
SED modelling. We use the height and shape of the warp from these analytic 
models as a guide for our dynamical models of the interaction of a disk with 
a dipole stellar magnetic field. We present the resulting photopolarimetry 
from this model, again using the disk structure derived from SED models for 
the outer disk.

We do not expect the time dependent 3D disk structure (due to shadowing of 
different regions of the outer disk by the warp) to effect our simulated 
{\it photometric} variability. The photometric variability is primarily 
produced by occultations of the star and the contribution from scattered 
starlight in the outer disk is small. However, the shape of the outer disk 
may be important for accurately modelling the {\it polarimetric} variability 
because it is produced by the scattering of starlight. We plan to explore 
three dimensional disk structure models in the future. In the meantime, we 
proceed with the modelling as outlined above and detailed below.

\subsection{Disk Structure from SED Modelling}

\begin{table}
 \caption{Dust Properties}
 \label{symbols}
 \begin{tabular}{@{}lcccccc}
  \hline
   & $\kappa$ (cm$^{-2}$g$^{-1}$) & $a$ & $g$ & $P$ (\%)&\\
  \hline
  $U$ &46.2 & 0.47  &0.64 & 39.3 \\ 
  $B$ &42.3 & 0.48  &0.63 & 41.1 \\ 
  $V$ &37.5 & 0.49  &0.62 & 40.6 \\ 
  $I$ &28.9 & 0.52  &0.60 & 38.1 \\ 
  \hline
 \end{tabular}
\end{table}

\begin{figure}
\psfig{figure=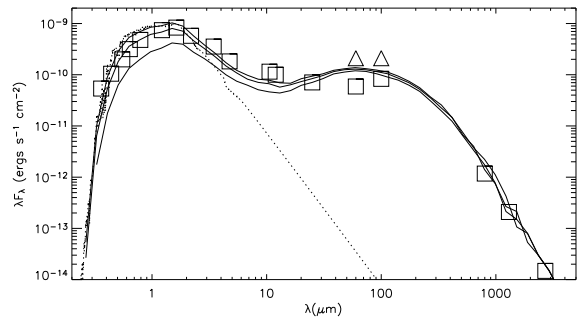,angle=90,width=8cm}
\caption{Spectral energy distribution data and model for AA Tau. The dotted 
line is the adopted input stellar spectrum and the three solid lines are the 
model SEDs for inclinations of (from top to bottom) $65^\circ$, $68^\circ$ 
and $71^\circ$. The different data points at 60 and 100~$\mu$m represent 
$IRAS$ (squares) and $ISO$ (triangles) observations.}
\end{figure}

AA~Tau's spectral energy distribution exhibits the large infrared 
excess emission characteristic of dusty protoplanetary disks. We model the 
SED with our Monte Carlo radiative equilibrium techniques 
(Bjorkman \& Wood 2001; Wood et al. 2002; Whitney et al. 2003) updated to 
include an iterative loop to calculate the disk structure for an irradiated 
steady accretion disk in vertical hydrostatic equilibrium 
(Walker et al. 2004). This approach follows D'Alessio et al. (1999) in 
adopting Shakara \& Sunyaev (1973) $\alpha$-disk theory to describe cTTS 
accretion disks.  Therefore the disk density is not parameterised by power 
laws as in our previous SED models (e.g., Wood et al. 2002; Rice et al. 2003).
In our simulations we adopt the dust opacity model which we have used to 
successfully model the SEDs of the HH~30~IRS and GM~Aur disks 
(Wood et al. 2002; Schneider et al. 2003; Rice et al. 2003). The 
wavelength dependence of this dust opacity is displayed in Wood et al. 
(2002, Fig.~2) and has $\kappa_B/\kappa_K=2.5$ which is in the range of 
``good fits'' as determined from the scattered light models of the HH~30~IRS 
disk using parametric disk models (Watson \& Stapelfeldt 2004).  The 
optical dust scattering properties (opacity, $\kappa$, albedo, $a$, 
phase function asymmetry parameter, $g$, and maximum polarisation, $P$) 
are shown in Table~1. These parameters are incorporated into our scattered 
light models as described in Code \& Whitney (1995).

Our iterative Monte Carlo radiative equilibrium code self-consistently 
calculates the disk density and temperature structure and emergent spectrum 
at a range of viewing angles.  From consideration of the photopolarimetric 
variability, AA~Tau's inclination has been estimated to be around $70^\circ$. 
Our simulations show that we can reproduce the AA~Tau SED at a viewing angle 
of $i\sim70^\circ$ with the following parameters $T_\star=4000$~K, 
$R_\star=1.9 R_\odot$, $R_d=150$~au, and $\dot M = 7.5\times 10^{-9}M_\odot 
{\rm yr}^{-1}$, corresponding to a total disk mass of $M_d=0.02M_\odot$.  
The code calculates an inner radius for the disk of $7R_\star$, corresponding 
to our adopted dust destruction temperature of 1600~K.   

Figure~1 shows observations of AA~Tau and our SED model for the system 
described above, viewed at $i=65^\circ$, $68^\circ$ and $71^\circ$. 
Observations come from the Kenyon \& Hartmann (1995) compilation (squares) 
with additional $ISO$ fluxes (triangles) from Chiang et al. (2001). The input 
stellar spectrum  is from a Kurucz model atmosphere (Kurucz 1994). 
For inclinations $i>70^\circ$, the direct optical starlight becomes obscured 
by the flared disk.  Therefore if the system inclination is indeed greater 
than $70^\circ$, this points to some potential shortcomings in our models.  
For example, some dust settling could effectively reduce the disk 
scale-height and therefore allow the star to be viewed at inclinations 
$i>75^\circ$ (e.g., Dullemond \& Dominik 2004). We have not attempted to fit 
the optical flux, since that is observed to vary and is the subject of the 
next section.  The point of this SED model is to obtain estimates for the 
disk mass and density structure based on a physically plausible disk model. 
We then use this as the outer disk structure in our subsequent optical 
scattered light models that use a warped inner disk.  Future work will 
investigate the effects of dust settling on the SED and inclination 
determinations.

\subsection{Photopolarimetry Models and Analytic Disk Warping}

The observed photopolarimetric variability in AA~Tau has been interpreted 
as eclipses of the star by a warp in the circumstellar disk.  To investigate 
this interpretation, we have constructed scattered light simulations where 
we introduce a warp into the axisymmetric disk geometry that we derived from 
SED fitting. Informed by the dynamical model (Section 4), where the disk 
midplane is not raised and lowered to produce the warp but material 
'piles up' at the corotation radius, we create the disk warp by introducing 
an azimuthal, $\phi$, dependence of the disk scale-height,
\begin{equation}
z_0=z_w \exp{ -{1\over 2} [(\phi-\phi_0)/\Delta\phi]^2  }
\exp{ -{1\over 2} [(\varpi-\varpi_0)/\Delta\varpi]^2  }
\end{equation}

\begin{figure}
\psfig{figure=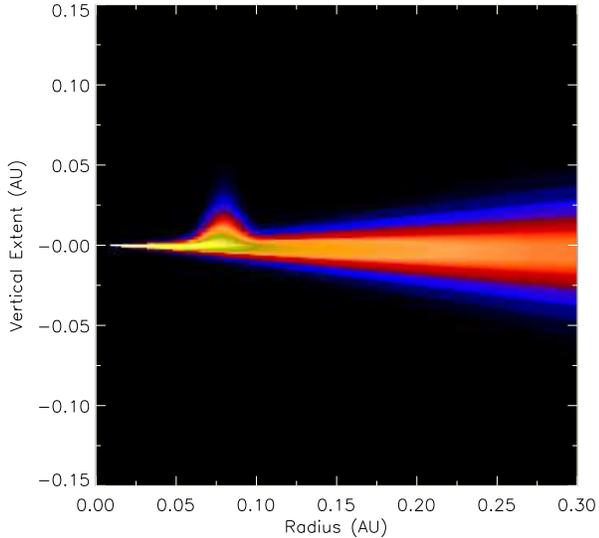,angle=0,width=8cm}
\caption{Analytically induced warp used to fit AA~Tau's photopolarimetric 
variability.  The figure shows a slice through the disk density at 
azimuthal angle $\phi=0^\circ$, showing the peak amplitude of the 
inner disk warp.}
\end{figure}

\begin{figure}
\psfig{figure=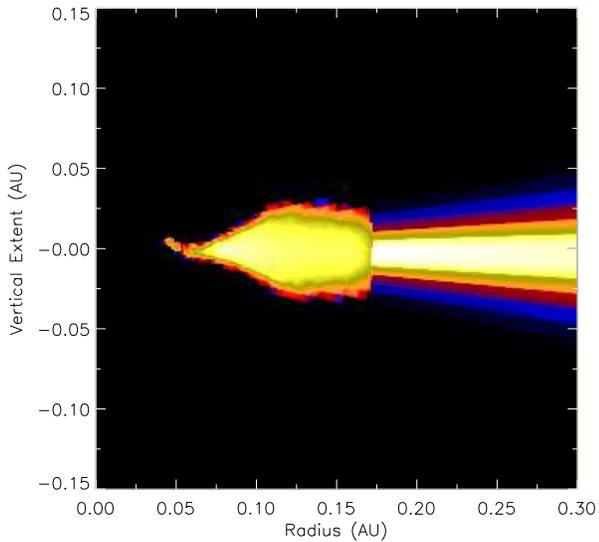,angle=0,width=8cm}
\caption{Dynamically induced warp used to fit AA~Tau's photopolarimetric 
variability. The figure shows a slice through the disk density, showing the 
peak amplitude of the disk warp.}
\end{figure}

with $z_w$ the amplitude of the warp, $\phi_0$ the azimuth of the warp, 
and $\Delta\phi$ the azimuthal extent of the warp. The warp is further 
constrained in radius with the second Gaussian function so that it peaks 
at $\varpi_0$ and extends over a radius $\Delta\varpi$. A warp is created 
in both the positive and negative $z$ directions, with the peaks having a 
$\pi$ phase separation. A second warp in the negative $z$ direction does not 
alter our simulated light curves for this viewing angle. Figure~2 shows the 
warp in the inner regions of the disk on the positive $z$ surface responsible 
for the photopolarimetric variability. Figure~3 shows the dynamically induced 
warp in the inner regions of the disk which is of similar dimensions but 
has a larger radial extent. As mentioned earlier the radial extent of the 
warp has little effect once it is large enough to fully obscure the central 
source so this makes no difference to the photopolarimetry of the model. 

\begin{figure}
\psfig{figure=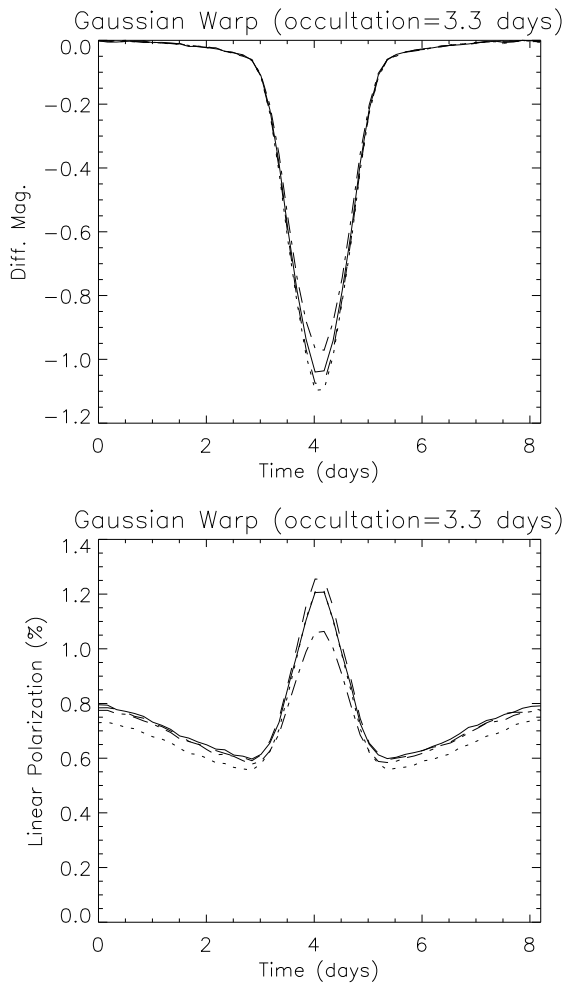,angle=0,width=8cm}
\caption{Photo-polarimetry simulations of a Gaussian shaped warp with an 
occultation duration of 3.3~days. The upper panel shows the variation in 
observed flux, with $\Delta m \sim 1$, for U (dotted),B (dashed),V (solid) 
\& I(dot-dashed) bands. The lower panel shows the corresponding linear 
polarisation with $P=0.4\% - 1.05\%$ in the V-band.}
\end{figure}

\begin{figure}
\psfig{figure=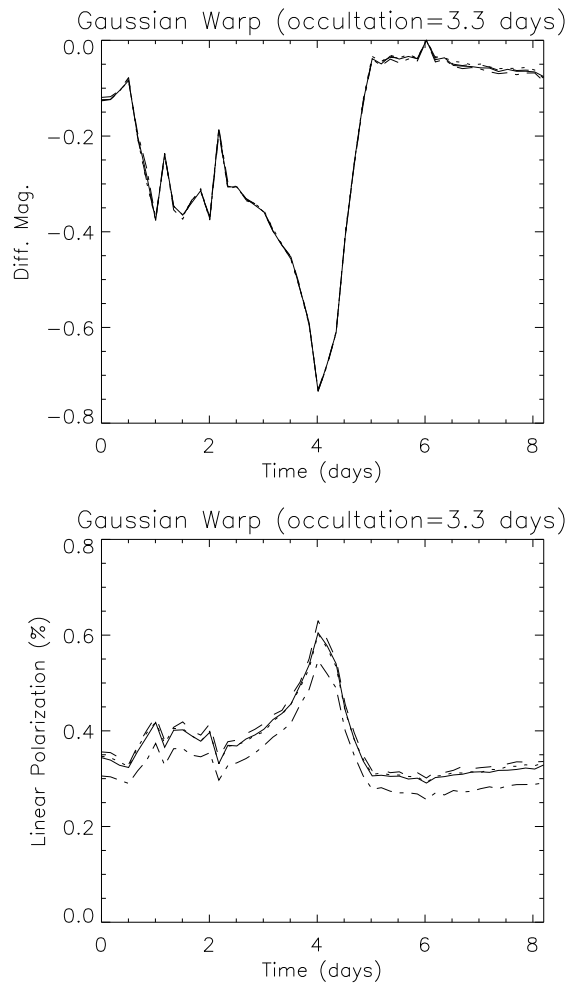,angle=0,width=8cm}
\caption{Photometry simulations of the dynamically induced warp with an 
occultation duration of $\sim 3$~days. The upper panel shows the variation in 
observed flux, with $\Delta m \sim 0.73~magnitude$, for U (dotted),B (dashed),
V (solid) \& I (dot-dashed) bands. The lower panel shows the corresponding 
linear polarisation with $P=0.6\% - 0.8\%$ in the V-band.}
\end{figure}

After the original interpretation that the variability of AA Tau 
was due to a warped inner disk (Bouvier et al. 1999), a study by 
Terquem \& Papaloizou (2000) examined the conditions under which an 
inclined stellar magnetic dipole could reproduce the required warp. They 
found that a dipole inclined at $30^\circ$ is easily capable of inducing a 
warp of the size proposed by Bouvier et al. (1999). They also discovered that 
depending on the viscosity of the disk the vertical displacement of the disk 
varied rapidly (low viscosity) and was likely to cause break up or caused a 
smoothly varying warp of the kind expected (high viscosity). Both cases would 
produce variations in light curves that could possibly be distinguished 
from one another.

\begin{figure*}
\psfig{figure=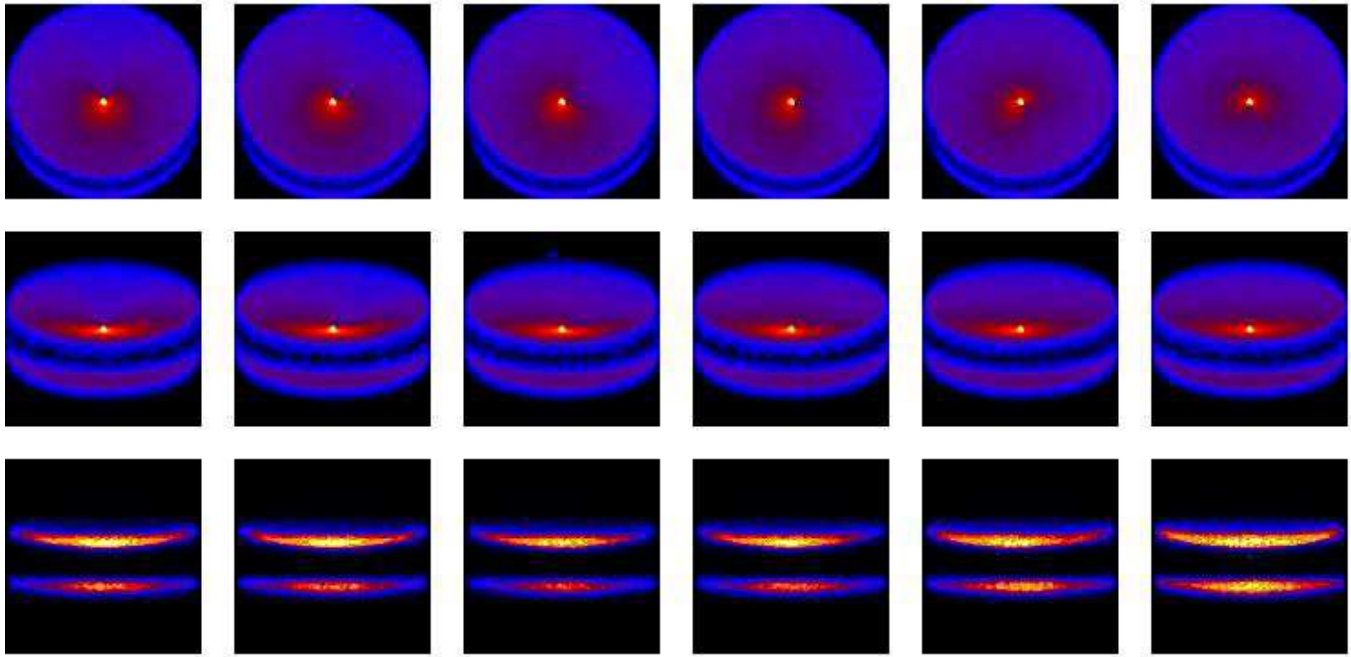,angle=90,width=16cm}
\caption{Scattered light images of our warped disk model of AA Tau. The 
panels on the upper row show the disk at an inclination of $25^\circ$, the 
centre row of panels have inclination $70^\circ$ and the panels along the 
bottom show an inclination of $85^\circ$, all are $400$~au on a side. The 
images cover half a rotation and clearly show the shadowed area caused by 
the warp occulting the star moving round the disk. To overcome the large 
dynamic range between starlight and scattered light in the disk, the images 
are presented on a one-tenth root (square root for the edge on disk) stretch. 
The faintest regions have a surface brightness of $10^{-6}$ that of the star.}
\end{figure*}

\begin{figure*}
\psfig{figure=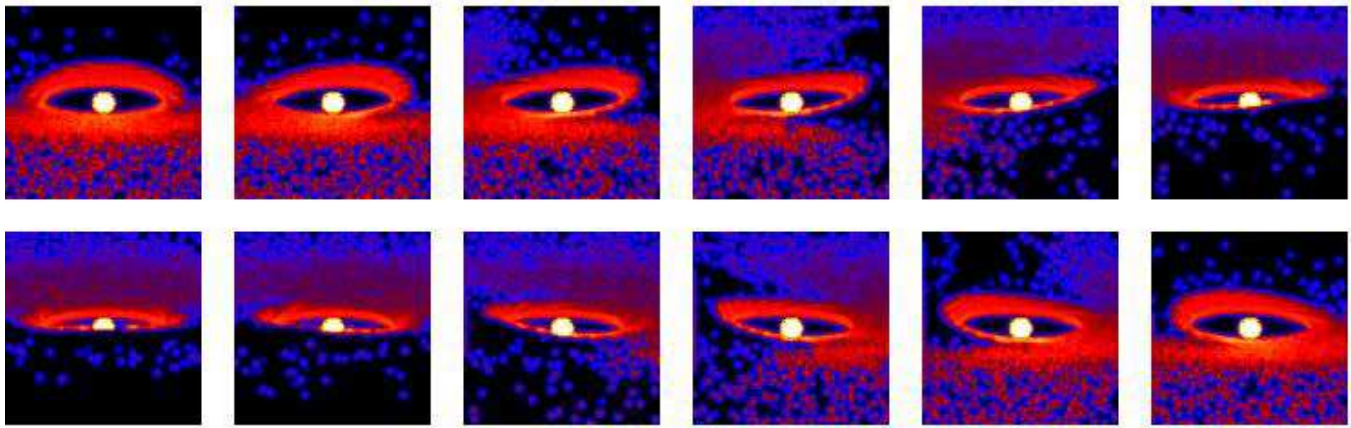,angle=90,width=16cm}
\caption{Scattered light images of the warped disk inclined at $75^\circ$ and 
with an image diameter of $0.2~au$ clearly showing the material responsible 
for the photopolarimetric variations occulting the star. To overcome the 
large dynamic range between starlight and scattered light in the disk, the 
images are again presented on a one-tenth root stretch. The faintest regions 
have a surface brightness of $10^{-6}$ that of the star.}
\end{figure*}

Our parameterised inner disk warp simulations that best reproduce the 
observed photopolarimetry have $i=75^\circ$, $z_w=0.016$, 
$\Delta\phi=75^\circ$, $\varpi_0=0.072$~au and $\Delta\varpi_0=0.0084$~au. 
Figure~4 shows the photpolarimetric light curves for this model. 
Note that the photometric variability for a warped disk is achromatic, as 
observed. Even though there is a wavelength dependence to the circumstellar 
dust opacity, the disk warp is optically thick enough that colour variations 
in the photometry are not observable. Slight wavelength dependent variations 
are present in the level of polarisation. In the observed polarimetry 
of AA Tau (Bouvier et al. 1999) there is a substantial position angle 
variation of around $0^\circ$ to $30^\circ$. By comparison with other stars 
in the vicinity a study by M\'{e}nard et al. (2003) attributed this 
variation to the interstellar medium. After ISM polarisation is removed 
$\Delta PA$ drops to $0^\circ$.In our simulations we find $\Delta PA$ 
$\sim 4^\circ$ indicating that there is in fact some intrinsic $\Delta PA$. 
If this is the case then it is not unreasonable to assume that the 
interstellar polarisation is $< 0.5\%$. The removal of a smaller interstellar 
polarisation component would in turn produce a slightly higher linear 
polarisation value for AA Tau, somewhere between the value stated by 
M\'{e}nard et al. (2003) and Bouvier et al. (1999) mentioned earlier.

Clearly many different warped disk models could match the observations, and 
we have constrained the parameters as follows:

The direct stellar flux and polarisation from a flared disk are 
sensitive to the inclination angle (e.g., Whitney \& Hartmann 1992).  
For viewing angles $i>78^\circ$, the star becomes blocked by the flared 
disk and the polarisation increases dramatically (e.g., see Stassun \& 
Wood 1999, Fig.~5) since the relative fraction of scattered to direct 
starlight increases.  For viewing angles $i<70^\circ$ we obtain very low 
polarisation values and would require very large warps, $z_w$, to obtain 
the observed photometric variability. We do not allow warps to exceed 
$z_w/\varpi_0 = 0.3$ in accordance with theoretical models of disk warping 
(Bouvier et al. 1999).  The warp must be constrained to a fairly narrow 
range in disk radius to obtain the observed variability. Warps extending 
over large radial distances would not survive due to differential rotation 
in the disk. The shape of the photopolarimetry light curve allows us to place 
constraints on the azimuthal extent of the warp. As the occultation itself 
is reported to last from 3-4 days it implies a warp with an azimuthal 
extent of $130^\circ-175^\circ$. The warp in figure~4 matches the observed 
brightness variation and polarisation of AA Tau. We find that as we increase 
the warp's extent around the disk there is very little change in the 
brightness variation but there is a reduction in the level of polarisation 
reached during the occultation. Values of $\Delta\phi$ much larger or smaller 
than what we use result in light curves that show too broad or too narrow an 
eclipse feature.

For comparison, modelling was carried out using a sinusoidal shaped warp.  
The duration of the warp was 3.3~days and there was very little change in the 
photopolarimetry compared to our previous model.For the sinusoidal warp, 
$\Delta V=1.08$ showing good agreement with the Gaussian model 
($\Delta V=1.05$) and the polarisation varied from $0.4\%-1.15\%$ also in 
good agreement with the Gaussian model where polarisation varied from 
$0.6\%-1.2\%$. 

Finally we modelled the photopolarimetry of the dynamical models 
magnetically induced warp. A stellar magnetic dipole of 5.2kG inclined at 
$30^\circ$ to the stellar rotation axis produced a warp of approximately the 
same dimensions as the analytical model. The duration of the occultation 
event was around 3.3 days with some low level variability lasting slightly 
longer. The warp produced a variation in photometry, $\Delta V$, of 0.73 Mag.
The polarisation was found to vary from $0.3\%-0.6\%$ giving good agreement 
with the analytical model and the observed variability. Figure~5 shows the 
photometric light curves for this model at various wavelengths. 

In summary we estimate the uncertainty in our models to be 
$i=75^\circ\pm 2^\circ$, $\Delta\phi=75^\circ\pm5^\circ$, 
$\Delta\varpi=0.0084 \pm 0.0042$~au and $z_w = 0.016 \pm 0.0016$~au.

\subsection{Time Sequence Scattered Light Images}

In addition to eclipsing the star and producing the observed unresolved 
photopolarimetric variability, the warp also casts a shadow over the 
outer regions of the disk.  Therefore, a time sequence of high 
spatial resolution images may detect a shadow sweeping round 
the disk. Figure~6 shows a sequence of scattered light images at a range of 
viewing angles for our AA~Tau warped disk model. Figure~7 shows a series of 
'close-up' scattered light images of the warp occulting the star at a viewing 
angle of $75^\circ$. These models may be compared with hot star spot models 
(Wood \& Whitney 1998) which show a lighthouse effect of a bright pattern 
sweeping around the disk.  For some warped disk models, the azimuthal extent 
of the warp may mimic scattered light images due to hot star spots, but multi 
wavelength photometry can discriminate models: star spots yield chromatic 
variability, whereas a disk warp yields achromatic photometric variability. 
Notice that for edge-on viewing the time sequence images for warped disks and 
disks illuminated by a spotted star (Wood \& Whitney 1998) are very similar, 
and multi-wavelength photopolarimetry is required to break the degeneracy.

For completeness we included in our warped disk models spots of various sizes 
located at a range of latitudes with a temperature of $8000$~K. 
We found that a spot with an angular radius of $5^\circ$ was more than enough 
to visibly alter the photometry at all latitudes causing the wavelength 
dependant effects mentioned above. Therefore a spot covering more than 
$0.2\%$ of the stars surface area causes a chromatic variation in the 
photopolarimetry that does not compare with observtions. Figure~8 shows 
unresolved photopolarimetric models of our warped disk illuminated by a star 
with hot spots on its surface. The spotted star model exhibits strong colour 
changes not present in the warped disk model. The shape of both the 
photometric variation and polarimetric variation curves are also quite 
different for the spot model purely as a function of the differing geometries 
involved. These strong colour changes are not reported in AA Tau, however, 
the HH~30~IRS disk does exhibit a colour dependence (Wood et al. 2000), so it 
appears that the hot spot models are more appropriate for that system.

\begin{figure}
\psfig{figure=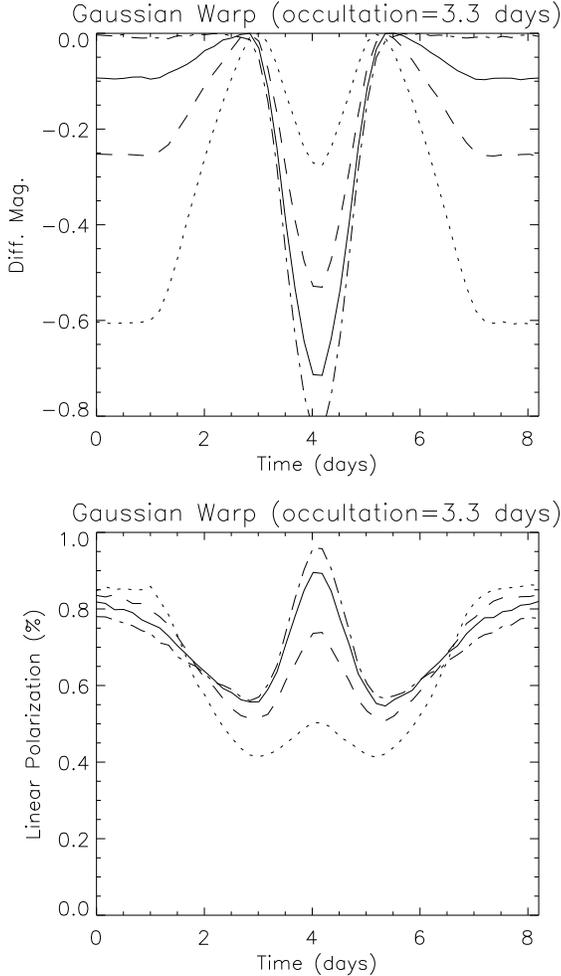,angle=0,width=8cm}
\caption{Variations in photometry (top panel) and polarisation (bottom panel) 
for our warped disk illuminated by a spotted star at 4 different 
passbands, U (dotted), B (dashed), V (solid) \& I (dot-dashed). The hot spot 
is at the same longitude as the warp, a latitude of $65^\circ$, has a radius 
of $5^\circ$ and a temperature of $8000$~K. Note the strong wavelength 
dependence and different shape of the light curve to that generated by a 
warped disk geometry.}
\end{figure}

\section{Hydrodynamic simulations}

Having analytically investigated the size and shape of the inner disk 
warping required to match AA~Tau's photopolarimetry, we now use a three 
dimensional hydrodynamics code to explore magnetically induced disk warps.  
We model the inner accretion disk region of AA Tauri with a three-dimensional
smoothed particle hydrodynamics (SPH) code. SPH is a Lagrangian numerical
scheme in which gas flow is represented by a system of particles moving with 
the local fluid velocity \citep{mon}. The SPH method has been applied 
successfully to accretion disks in a host of astrophysical situations 
including protoplanetary disks \citep{ric}, cataclysmic variables 
\citep{tru00} and micro-quasars \citep{tru04} and has been applied to the 
magnetic warping of disks by \citet{mur}. The warping of a disk in response 
to an offset dipolar field has also been calculated with a three-dimensional 
Eulerian magnetohydrodynamics code by \citet{rom}.

In our model, we use operator-splitting to solve for the dynamics of the gas 
flow subject to three forces. The gas pressure force is computed by solving 
the SPH momentum equation with the standard SPH viscosity term:
\begin{equation}
   \frac{d{\bf v}_i}{dt}=-\sum_j m_j \left( \frac{P_i}{\rho_i^2}+\frac{P_j}
{\rho_j^2}+\frac{\beta \mu_{ij}^2-\alpha \bar c_{ij} \mu_{ij}}
{\bar \rho_{ij}} \right) \nabla_i W_{ij}.
\label{momeq}
\end{equation}
Here, $W_{ij}$ is the interpolating kernel between particles i and j, 
$\bar c_{ij}$ is the mean sound speed of the two particles and $\mu_{ij}$ is 
such that
\begin{equation}
 \mu_{ij} = \left \{ \begin{array}{ll}
\frac{H{\bf v}_{ij} \cdot {\bf r}_{ij}}{{\bf r}_{ij}^2 + 0.01h^2} & \mbox{$
{\bf v}_{ij} \cdot {\bf r}_{ij} \leq 0$} \\
0 & \mbox{${\bf v}_{ij} \cdot {\bf r}_{ij} > 0$}
\end{array}         \right .
\label{piv}
\end{equation}
where $h$ is the smoothing length and $H$ is the local scale height of the 
disk. The viscosity parameter $\alpha$, should not be confused with the 
Shakura-Sunyaev viscosity parameter, although \citet{mur96} has shown that in 
three dimensions with $\beta$=0, the net Shakura-Sunyaev viscosity introduced 
by this model is
\begin{equation}
  \alpha = \frac{1}{10} \alpha_{\rmn{SPH}}.
\end{equation}

The gravitational attraction of the star is computed via a simple Runge-Kutta 
fourth order integrator. We do not consider the self-gravity of the accretion 
disk, as we are modelling only a small, low-mass region near the central star.
Full MHD is not yet possible with SPH, so a third force is added, representing 
the drag on each particle due to a magnetic dipole field anchored on the star.
The dipole field is assumed to co-rotate with the star, but is offset slightly 
with the rotational axis.

\begin{figure}
\psfig{file=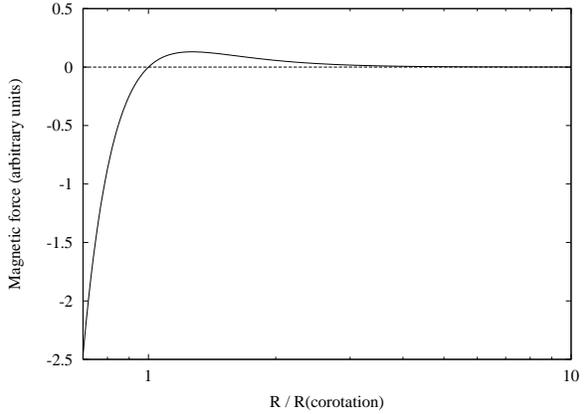,width=8cm,angle=-90}
\caption{The radial dependence of the magnitude of the magnetic drag 
acceleration term, plotted here for a surface density profile $\Sigma \sim 
\rho H \propto 1 / r$. In the simulations, the surface density is calculated 
self-consistently, but in practise departures from this profile are small. 
The net drag force is directed toward the central star inside the co-rotation 
radius and its magnitude increases sharply towards the star. Outside the 
co-rotation radius, the net drag force is weaker and is directed away from 
the central star. The drag force is zero at $R = R_{\rmn{co}}$.}
\label{mag}
\end{figure}

The magnetic drag force model has been described by \citet{wyn}, and was 
first included in a SPH scheme by \citet{mur} to investigate the magnetic 
warping of disks in cataclysmic variables. The model has been developed 
further in a recent paper by \citet{mat}, in which it is used  in a 
one-dimensional model of accretion disks in T Tauri stars. Here, we 
incorporate these developments into a fully three-dimensional hydrodynamic 
study of a circumstellar disk. In the model, the magnetic tension force 
appears as 
\begin{equation}
  a_{\rmn{mag}} \sim \frac{B_{\rmn z}^2}{4 \pi \rho r_{\rmn{c}}} \left( \frac
{\Omega - \Omega_*}{\Omega} \right),
\end{equation}
where $\Omega$ and $\Omega_*$ are the angular velocities of the gas and the 
star respectively and $r_{\rmn{c}}$ is the local radius of curvature of the 
magnetic field lines. This is approximated as a fraction of the local scale 
height of the disk, 
\begin{equation}
  r_{\rmn{c}} = \zeta H
\end{equation}
where $\zeta \leq 1$ \citep{pea}.

For a dipole of magnetic moment $\mu$ we have
\begin{equation}
 B_{\rmn z} = \frac{\mu}{r^3}.
\end{equation}
The drag force acts in a direction perpendicular to the relative velocity of 
the gas and the rotating field. It is positive, propelling material away from 
the star, for all radii $r > R_{\rmn{co}}$, where $R_{\rmn{co}}$ is the 
co-rotation radius. Conversely, gas at radii $r < R_{\rmn {co}}$ feels a net 
force towards the star. The functional form of this force term is plotted in 
Figure \ref{mag}. For numerical convenience, we only model the flow of gas 
outside a radius $R_{\rmn{min}} = 4 R_*$, since close to the star the magnetic
drag force becomes very large. This has no impact whatsoever on the resolution
of structure in the accretion disk itself, which is truncated well outside 
this radius.

\begin{figure*}
~~~~~~\psfig{file=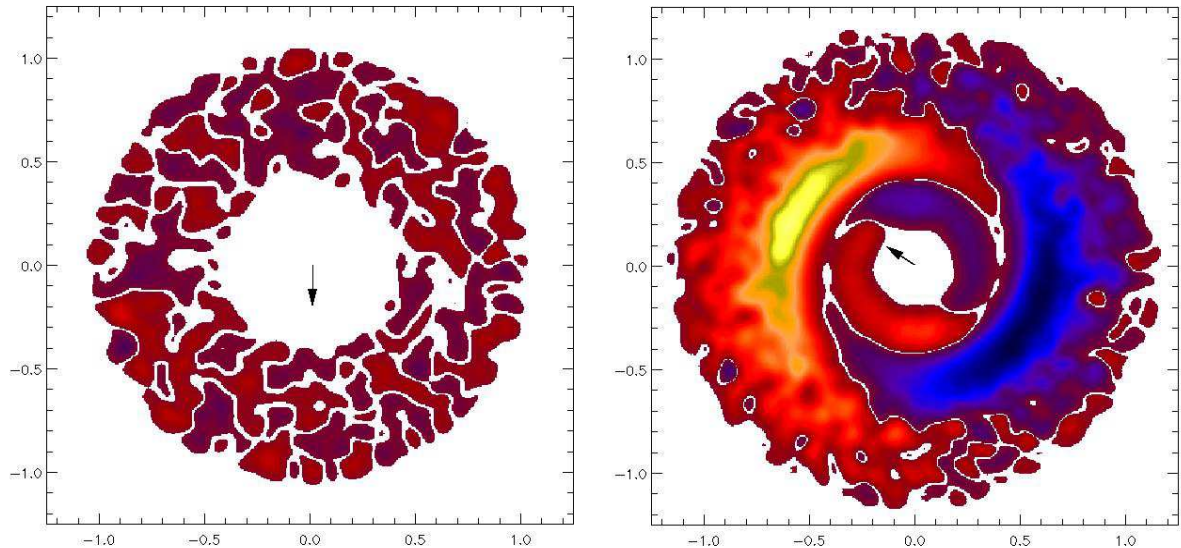,width=16cm}
\caption{These images of the accretion disk viewed from directly above the 
star are coloured according to the mean height from the mid-plane at the 
beginning (left) and end (right) of the simulation. Blue regions are, on 
average, below the mid-plane while yellow and orange are, on average, above 
the mid-plane. Regions that remain coplanar with the stellar equator are left 
white. The initial conditions (left) are chosen such that the altitude of gas 
at each radius is symmetric about the mid-plane, hence the mean height appears 
random (but is very close to zero). The axes are scaled in units of $20 R_*$. 
Gas that appears inside the co-rotation radius (which lies near 0.45 in these 
units) is threaded onto the magnetic field lines of the stellar dipole field.
The black arrows at the centre of the disk show the projected direction of the
north pole of the dipole, which is offset at an angle $30^{\,\circ}$ to the 
vertical.}
\label{maps}
\end{figure*}

We set up an initial disk comprising 500000 SPH particles, extending from the 
corotation radius $R_{\rmn{co}} = 8.7 R_*$ to $20 R_*$, where the stellar 
radius is taken to be $R_* = 1.3 \times 10^{11} {\rmn{cm}}$. The midplane of 
the initial disk is coplanar with the stellar equator and has a surface 
density profile $\Sigma(r) \propto r^{-1}$. It is flared, with a hydrostatic 
vertical density profile $\rho(z) = \rho(0)\,{\rm exp}(-z^2 / 2 H^2)$, where 
$\rho(0)$ is the density in the mid-plane and $H$ is the scale height 
$c_{\rmn s} / \Omega$. The particles are given radial velocities such that 
throughout the disk the mass accretion rate is constant at a value 
$\dot M = 7.5 \times 10^{-9} M_\odot {\rm yr}^{-1}$.Here, we are considering 
the case of a constant mass accretion rate into the inner regions of the 
accretion disk. Naturally, it is quite possible to surmise that any variations
in mass transfer rate will impact on the size and nature of the warping of the
disk. This scenario is discussed in a recent paper by \citet{pfe}. We comment 
that another factor that may affect the long-term behaviour of any warp is 
magnetic diffusivity, which we do not consider here. It seems likely that a 
finite resistivity will modify the field structure somewhat, although the 
importance of this effect for the long-term stability of a disk warp remains 
unclear.

We use a Shakura-Sunyaev viscosity parameter $\alpha = 0.01$, $\beta = 0$ and 
a dipole moment $\mu =1.2 \times 10^{37}\,{\rmn{Gcm^3}}$. With $\zeta = 1$, 
this corresponds to a stellar field strength $B(R_*) = 5.2 {\rmn{kG}}$, which 
is slightly larger than the values usually quoted for T-Tauri stars, in the 
range $2 -3 {\rmn{kG}}$. This field was required to produce the warp described
below for the parameters given, although it should be stressed that there is 
nothing to prevent $\zeta < 1$, in which case there is better agreement with 
the observed field strengths. The magnetic dipole is offset at angle $\pi / 6$
to the axis of stellar rotation.

After several orbits under the influence of the magnetic field, a stable 
warped structure develops near the corotation radius. Figure \ref{maps} shows 
the local average height of the disk above the mid-plane, in the initial and 
final states. The warp has a maximum vertical height of $ \sim 2 R_*$. This 
result is consistent with the earlier analysis of \citet{ter}, who computed 
the steady warped structure of the disk in AA Tauri, and predicted a similar 
trailing spiral structure near the corotation radius. We also made a limited 
study of the effects of changing the tilt angle of the magnetic dipole. 
Increasing the tilt angle to $5\pi /18$ had little or no effect on the size of
the warp, and the resultant disk structure was indistinguishable from that 
obtained in the original simulation.

\section{Summary}

We have modelled the photopolarimetric variations of the classical T Tauri 
system AA Tau. Our results show that a magnetospherically induced warp of 
the accretion disk at roughly the stellar corotation radius occults the star 
and reproduces the observed variability. Our SED modelling provides us with 
estimates of the disk mass and large scale density structure that are 
subsequently used in our non-axisymmetric scattered light disk models. 
Spotted star models exhibit a strong wavelength dependence which is not 
observed in the AA Tau system. Our warped disk model shows no wavelength 
dependence and can reproduce the occultation period and duration with the 
required brightness and polarisation variations. A feature of the warped disk 
model is that it produces a shadow that sweeps around the outer disk and this 
may be detectable with high spatial resolution time sequence imaging. 

Using a modified SPH code, we find that a stellar magnetic dipole 
of 5.2kG inclined at $30^\circ$ to the stellar rotation axis may reproduce 
the required warp amplitude to occult the star and reproduce the 
brightness variations. The models we have presented are strictly periodic, so 
do not reproduce the stochastic nature of AA~Tau's lightcurve 
(Bouvier et al. 2003). However, our models do show that disk warping 
resulting from the interaction of the stellar magnetic field with the disk 
can reproduce the amplitude and shape of the occultation events. In the near 
future, accurate measurements of the magnetic field structures of T Tauri 
stars will be possible using Zeeman Doppler Imaging (Petit et al. 2004), 
allowing more realistic (i.e., non-dipolar) modelling of the stellar 
magnetic field, its impact on the disk, and observational signatures.

We acknowledge financial support from PPARC studentships (MO, CW, OM), 
a Postdoctoral Fellowship (MRT), and an Advanced Fellowship (KW).

\label{lastpage}

\end{document}